\documentstyle[twocolumn,amssymb,aps,psfig,float]{revtex}
\begin{document}

\newcommand{\bsigma}{\mbox{\boldmath $\sigma$ } }
\newcommand{\bS}{\mbox{\boldmath $S$ } }
\newcommand{\bT}{\mbox{\boldmath $T$ } }
\newcommand{\bLambda}{\mbox{\boldmath $\Lambda$ } }
\newcommand{\bmu}{\mbox{\boldmath $\mu$ } }
\newcommand{\bk}{\mbox{\boldmath $k$ } }
\newcommand{\bR}{\mbox{\boldmath $R$ } }

\twocolumn[\hsize\textwidth\columnwidth\hsize\csname @twocolumnfalse\endcsname

\title{
Dynamic spin-glass behavior in a disorder-free, two-component model
of quantum frustrated magnets }
\vskip0.5truecm
\author{Fr\'ed\'eric Mila$^{(a)}$ and David Dean$^{(b)}$}
\address{
$^{(a)}$ Institut de Physique Th\'eorique, Universit\'e de Lausanne,
CH-1015 Lausanne, Switzerland\\
$^{(b)}$ Laboratoire de Physique Quantique, Universit\'e Paul Sabatier, 31062
Toulouse, France}
\vskip0.5truecm

\maketitle

\begin{abstract}
\begin{center}
\parbox{14cm}
{ Motivated by the observation of a spin-glass transition in 
almost disorder-free Kagome antiferromagnets, and by the specific
form of the effective low-energy model of the S=1/2, trimerized
Kagome antiferromagnet, we investigate the possibility to obtain
a spin-glass behavior in two-component, disorder-free models.
We concentrate on a toy-model, a modified Ashkin-Teller 
model in a magnetic field that couples only to one species of spins,
for which we prove that a dynamic spin-glass behavior occurs.
The dynamics of the magnetization is closely related to that of
the underlying Ising model in zero field in which spins and pseudo-spins are
intimately coupled. The spin-glass like history
dependence of the magnetization is a consequence of
the ageing of the underlying Ising model.}
\end{center}
\end{abstract}
\vskip .1truein


\vskip2pc
]

\section{Introduction}

The experimental investigation of quantum magnets is undergoing rapid
progress with the synthesis of new and better controlled samples. Of
special interest are  frustrated magnets, for which very unusual
behavior has been reported\cite{ramirez}. In particular, a behavior 
reminiscent
of spin glasses has been reported in a number of Kagome antiferromagnets (AF)
\cite{ramirez,wills}.
While the role of disorder in these phenomena cannot be excluded yet, the
persistence of this behavior in progressively cleaner samples calls for
explanations in terms of disorder-free models.

The possibility of a spin-glass behavior in the spin 1/2 Heisenberg
model on the Kagome lattice without any disorder was first suggested in
the pioneering work of
Chandra et al\cite{chandra1} in 1993. 
Since the classical Heisenberg model on the Kagome lattice does not 
exhibit a spin-glass behavior, the main problem one is facing is how 
to include the quantum aspects of the 
problem into a description in terms of classical variables for which
one can use standard techniques to study the dynamics.
In Ref.\cite{chandra1}, the authors assumed that quantum fluctuations
select coplanar configurations, which led them to concentrate on the
anisotropic XY version of the model. For this model, they suggested
the presence at finite temperature of 3-spin order, and of a glassy transition
related to the binding of non-abelian defects that
are expected to be the natural point defects associated with this type
of order. For details, the reader should consult Ref.\cite{chandra1}. 
To prove or disprove this scenario turned out to be difficult,
and whether this indeed provides an explanation of
the spin-glass behavior of Kagome
antiferromagnets without introducing any disorder remains unsettled.
Since then, clear evidence of glassiness has been reported by Chandra
et al\cite{chandra2} for another class of disorder-free models describing
periodic Josephson arrays in a transverse magnetic field, but these models are
not directly related to the Kagome AF.

In parallel, a lot of progress has been made
in the understanding of the spectrum of the S=1/2 AF Heisenberg model 
on various
frustrated lattices, in particular the Kagome and the 
pyrochlore ones\cite{ramirez}.
So far, it is well established that frustration can have 
two effects: It can open a
gap to triplet excitations\cite{majumdar}, like for the non-frustrated
spin 1 chain\cite{haldane}, but it can also lead to a proliferation
of low-lying singlets inside this gap, like for the S=1/2 Kagome
antiferromagnet\cite{lecheminant}. These singlets can be interpreted as
RVB (Resonating Valence Bond) states\cite{elser,zeng,mambrini}, and they
could lead to a power-law behavior of the low-temperature specific
heat\cite{sindzingre,ramirez2,bernu}.
Experimental
systems known so far have a larger spin however (3/2, 5/2,...)\cite{ramirez},
and there is room for new physics in these cases since the presence of a
singlet-triplet gap is unlikely given the rather small value already
reported for S=1/2. Possible implications of this strange spectrum regarding
in particular a possible spin-glass behavior have not been discussed yet,
mainly due to the lack of methods to attack this problem.

\begin{figure}[hp]
\centerline{\psfig{figure=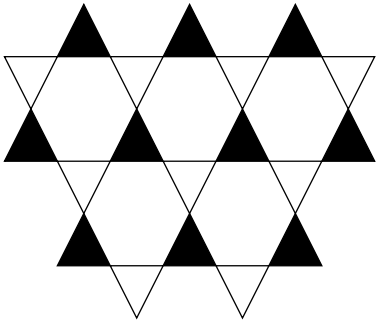,width=5.0cm,angle=0}}
\vspace{0.5cm}
\caption{Trimerized Kagome lattice showing the difference between
strong bonds (black triangles) and weak bonds (white triangles).}
\label{fig0}
\end{figure}

In this paper, we continue the quest initiated
in Ref.\cite{chandra1} for spin-glass behavior in the disorder-free, 
quantum Heisenberg model on the Kagome lattice. 
However, building on the recent results obtained
on the low-energy spectrum of the model, we propose another
approximate way of including quantum fluctuations into a classical
description. The starting point is the effective model obtained
in Ref.\cite{mila} for the spin 1/2 Heisenberg model on the trimerized 
Kagome lattice (see Fig. (\ref{fig0})). This is a modified version of 
the spin 1/2 Heisenberg model on the Kagome lattice in which the exchange
integrals take two different values $J$ and $J'$ according to the 
pattern of Fig. (\ref{fig0}). This is actually
the relevant description of the Kagome layers in 
SrCr$_{9p}$Ga$_{12-9p}$O$_{19}$ since the presence of a triangular
layer between pairs of Kagome layers lead to two types of bonds with 
precisely the pattern of Fig. (\ref{fig0}). Then, since the ground
state of a triangle is fourfold degenerate and can be described
by two spin 1/2 degrees of freedom, the total spin $\vec \sigma$ and 
the chirality pseudospin $\vec \tau$, it was shown in Ref.\cite{mila} 
that the low-energy effective Hamiltonian in the limit $J'\ll J$ 
can be written
\begin{eqnarray}
\tilde H & = & (J'/9) \sum_{<i,j>} \tilde H^\sigma_{ij} \tilde H^\tau_{ij},\ \ 
\tilde H^\sigma_{ij}  = \sum_{<i,j>} \vec \sigma_i.\vec \sigma_j,
\nonumber \\
\tilde H^\tau_{ij} & = & 
(1-2(\alpha_{ij}\tau_i^- +
\alpha_{ij}^2\tau_i^+))(1-2(\beta_{ij}\tau_j^- + \beta_{ij}^2\tau_j^+))
\label{effective}
\end{eqnarray}
where $<i,j>$ denotes pairs of nearest neighbors.
In $\tilde H^\tau_{ij}$, $\alpha_{ij}$ and $\beta_{ij}$ are complex parameters
that take the values 1, $\exp(2\pi i/3)$ or $\exp(4\pi i/3)$ depending
on the bond (for details, see Ref.\cite{mila}).

In the case of the trimerized Kagome lattice, quantum fluctuations
thus introduce an extra local degree of freedom in the Hamiltonian.
For the general case, the derivation is no longer valid since
$J'/J$ is not a small parameter, but the low-energy
spectrum found in exact diagonalizations of small clusters\cite{lecheminant}
is still consistent with such a description: The exponentially large
number of low-lying singlets suggest that there must be another quantum
number in addition to the spin to classify these eigenstates. This
is precisely the role played by the chirality pseudo-spin in the 
trimerized Kagome lattice. A detailed comparison of both spectra
has actually shown that they are very similar indeed\cite{mambrini}.

We would therefore like to propose a new approach to the spin-glass
behavior of disorder-free quantum magnets in terms of models
with two-local degrees of freedom. Our basic motivation for  concentrating on
such models is that they are {\it a priori} good candidates to exhibit
spin-glass behavior even without any disorder in the Hamiltonian. The
basic argument is the following. In a field-cooled (FC) experiment, in which
one quenches the sample from high temperature in a magnetic field,
the system will
automatically choose configurations in which the spins are polarized,
and the pseudo-spins will adapt to keep the energy as low as
possible. Hence the resulting magnetization at low temperature is
expected to be significant. However,
in a zero-field cooled (ZFC) experiment, the system is first quenched
without a magnetic field, and it will
use both degrees of freedom
to minimize the energy. When the magnetic field is switched on, in order to
polarize, the system will have to overcome the intimate mixing of spins
and pseudo-spins to get a significant polarization. This is likely to
take a very long time, and the apparent susceptibility will be much
smaller than the one measured in a FC experiment. 

The analysis of the model of Eq. (\ref{effective}) with the spin
and the pseudo-spin treated as classical Heisenberg variables is
a considerable task, and before starting such an endeavor, one
would like to know whether the scenario outlined in the previous
paragraph can indeed lead to a spin glass behavior. In fact, other
models, like the fully frustrated XY model\cite{villain} or some
vector models\cite{kawamura}, can be described
in terms of spin and chirality variables, and no spin-glass
behavior was ever reported for these models.
Consequently the rest of the paper is devoted to a detailed analysis 
of a toy model
to test whether the presence of two local degrees of freedom
can indeed lead to a spin-glass behavior.
The simplest model of this kind is a modified Ashkin-Teller model
defined by the Hamiltonian:
\begin{equation}
{\cal H}=J \sum_{\langle i,j\rangle} S_i S_j T_i T_j - h\sum_i S_i
\label{model}
\end{equation}
In this Hamiltonian, $S_i$ and $T_i$ are Ising variables that describe
the spin and the pseudo-spin respectively. Note that the magnetic
field $h$ is coupled only to the spin degree of freedom. Since
frustration has already been included in the model as an extra degree
of freedom, there is no need to work on a frustrated lattice any more,
and for simplicity we study this model on a square lattice. This
model is of course a very dramatic simplification since the Heisenberg
spins and pseudo-spins are replaced by Ising variables. Still, as
we shall see, the physics of this model is very rich, and to a
large extent confirms the simple picture of the previous paragraph.
Shortcomings that might be overcome by going to the more physical
description in terms of Heisenberg spins will be discussed in the last section.

\section{Monte-Carlo results}

Let us first discuss the equilibrium properties of this model. If
$h=0$, this model is equivalent to the antiferromagnetic Ising model
after the local gauge transformation $\sigma_i\equiv S_iT_i$,
and all states have an additional
degeneracy of $2^{N_{sites}}$ since $\sigma_i=1$ (resp. $-1$) can be
achieved with $(S_i,T_i)=(1,1)$ and $(-1,-1)$ (resp. $(1,-1)$ and
$(-1,1)$).  In fact, for any value of $h$, the partition function can
be factorized as $Z=Z_{\rm Ising}Z_S $
where $Z_{\rm Ising}$ is the partition function of the Ising model on a
square lattice, and $Z_S$ is the partition function of paramagnetic
spins in a magnetic field.  Accordingly, the free energy per site $f$
is given by:
\begin{equation}
f=f_{\rm Ising}-k_BT \ln (2\cosh(\beta h)) \label{elibre}
\end{equation}
with $\beta=1/k_BT$.
The magnetization per site is defined as usually by $m=-\partial
f/\partial h$, and since $f_{Ising}$ does not depend on $h$, we
obtain $m=\tanh(\beta h)$. So if the system has reached its
equilibrium state, the magnetization smoothly saturates as  the
temperature goes to zero.

However, as we shall see below, this equilibrium state might be very
difficult to reach depending on the history of the system.
To be specific, let us consider the following protocol: The quench
takes place at $t=0$, the field is switched on after a waiting time
$t_w$, and the magnetization is measured after a measuring time $t_m$
elapsed after the field has been switched on. The time elapsed between the
quench and the measurement is denoted by $t$. In a FC experiment, $t_w=0$  and
$t=t_m$, while in a ZFC one, $t_w>0$ and $t=t_w+t_m$.

\begin{figure}[hp]
\centerline{\psfig{figure=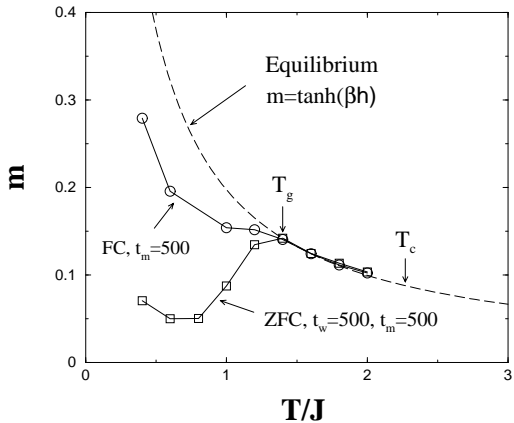,width=7.0cm,angle=0}}
\vspace{0.5cm}
\caption{Temperature dependence of the magnetization for different measurement
conditions.}
\label{fig1}
\end{figure}

To mimic such experiments, we have performed Monte Carlo simulations
of the model of Eq. (\ref{model}). The elementary step consists in
flipping either a spin or a pseudo-spin according to Glauber prescription.
The site and the variable (spin or pseudo spin) are chosen randomly, and
the time unit corresponds to a number of steps equal to the number of
degrees of freedom, i.e. twice the number of sites. In all numerical
experiments reported below, the starting configuration is completely
random, corresponding to an infinite temperature, and when it is not switched off,
the magnetic field $h$ is equal to 0.2 in units of $J$. Typical results
for the magnetization as a function of temperature are shown in
Fig. (\ref{fig1}) for a system of size 400 $\times$ 400.
Below a temperature $T_g$ which is smaller than the Curie
temperature $T_c\simeq 2.27 J$ of the underlying Ising model,
there is a clear difference between the ZFC and FC measurements: The ZFC
magnetization drops quite
abruptly, as in typical spin-glasses. Note that these curves depend
on $t_m$ and $t_w$ (this dependence should of course vanish if $t_m$
was infinite) but only weakly in a large parameter range, as we shall explain
below.

\begin{figure}[hp]
\centerline{\psfig{figure=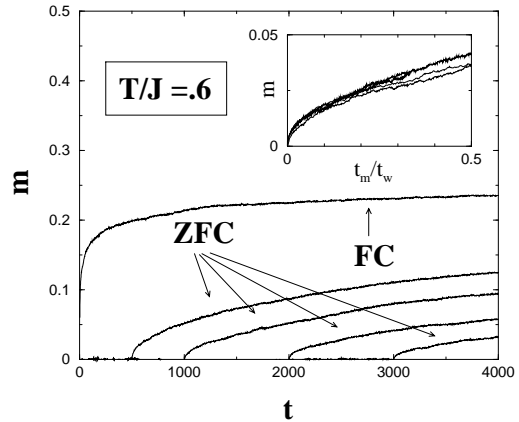,width=7.0cm,angle=0}}
\vspace{0.5cm}
\caption{Time dependence of the magnetization at given temperature for
various measurement conditions (FC and ZFC with $t_w=500, 1000, 2000, 3000$).
Inset: Plot of the ZFC data as a function of $t_m/t_w$.}
\label{fig2}
\end{figure}

To gain some insight into the origin of this behavior, it is very useful
to study the time evolution of the magnetization at fixed temperature
for different values of $t_w$. Typical results are given in Fig. (\ref{fig2}).
The most salient features are: i) A much faster increase at short times
for the FC experiment ($t_w=0$) than for the ZFC ones; ii) A similarity of the
shape of the ZFC curves. In fact, a very good scaling can be
obtained if we plot the magnetization as a function of $t_m/t_w$ (see inset of Fig.
(\ref{fig2})). Such a scaling is typical of ageing phenomena
\cite{BCKM}.

\section{Analytical approach: ageing and persistence}

To understand the behavior of the ZFC and FC
experiments let us first consider the
dynamics of the underlying Ising model in the absence of a magnetic field.
After
a quench below the Curie temperature $T_c$, the model has a spontaneous
staggered magnetization in the variables $\sigma_i$ in terms of statics.
However, in the absence
of a field that breaks the symmetry between the two ground states,
no global magnetization develops for an infinite system and the
system stays out of equilibrium. Domains of the two phases form and coarsen
with a characteristic length scale $\sqrt{t}$
\cite{bray} as the  evolution is via diffusion of
the domain walls and coalescence of domains.
The ergodic time $t_{\rm erg}$ for such a system scales with the
linear size $L$ of the system
like $L^2$, and
thus as long as $t \ll t_{\rm erg}$, the system is out of equilibrium and
has no global magnetization. We have checked that for  $T=.6 J$ and $L=400$,
the ergodic time is of the order
of 4000. Since this regime is the only one accessible for very large samples,
we will limit our discussion to that regime.


Another point worth noticing about the dynamics of the Ising model is that,
after the short initial stage where spins with 3 or 4 parallel neighbors
have disappeared, flips take place only at sites
with two neighbors up and two neighbors down. For such sites,
the effective field coming from the coupling to the neighbors
is equal to zero, and we call them 0-sites. Since there is no energy cost
to flip the spin $\sigma_i$ of such a site,
the probability to flip according to Glauber dynamics is 1/2. So the
flipping rate should be half the concentration of 0-sites. We have checked
numerically that this is indeed the case after a few sweeps. For $T=.6 J$, this
is true as soon as $t>40$.

Let us now turn to an analysis of the problem in the presence of a magnetic
field. Let us forget the first few steps and concentrate on
times where flips take place at 0-sites. If the magnetic field is small
compared to $J$, a condition usually fulfilled in experiments,
the dynamics of the Ising spins $\sigma_i$ will be essentially unaffected, and
the magnetic field will just favor configurations where the spins $S_i$ are
parallel to the field\cite{note}.
Now let us suppose that a 0-site remains so for
a while. The equation that governs the appearance of a magnetization for the
$S$ spins can be easily deduced. If we denote by $p_+$ (resp. $p_-$) the
probability for $S$ to be up (resp. down), Glauber dynamics implies that
$p_+$ satisfies the equation
\begin{equation}
\frac{dp_+}{dt}=\frac{e^{\beta h}}{e^{-\beta h}+ e^{\beta h}}\ p_-\  -
\frac{e^{-\beta h}}{e^{-\beta h}+ e^{\beta h}}\ p_+
\end{equation}
The magnetization, which is related to $p_+$ and $p_-$ by $p_+=(1+m)/2$ and
$p_-=(1-m)/2$, is thus given by
\begin{equation}
m=\tanh(\beta h)(1-\exp(-t/\tau))\end{equation}
where the relaxation time $\tau$ is equal to 1 in the chosen time units. This
is a very short time, especially considering the very small flipping rate
which is achieved after a few steps (already below $10^{-2}$ after 100 sweeps
for  $T=.6 J$).
Under these circumstances, the sites
that have been 0-sites in the presence of the magnetic field should have
enough time to reach equilibrium, and on average their magnetization
should be equal to $\tanh(\beta h)$. So if we call
$c_0(t_m,t_w)$ the proportion of sites that have been 0-sites between $t_w$
and $t_w+t_m$  this simple argument leads to the prediction that
\begin{equation}
m(t_m,t_w)=\tanh(\beta h) c_0(t_m,t_w) \label{gen}
\end{equation}
Note that $c_0(t_m,t_w)$ depends only on the dynamics of the underlying Ising
model and {\it not} on the magnetic field. To check this prediction, we
have calculated $c_0(t_m,t_w)$ for different $t_w$ corresponding to our
ZFC numerical experiments. The agreement is very good -
the curves are indistinguishable from the the ZFC calculations of $m$
on the scale of Fig. (\ref{fig2}) - and we have checked that
it remains so as long as $t_w$ is not too small so that the magnetization
is indeed controlled by 0-sites. This analysis shows that the ageing behavior
of the model of Eq. (1) is indeed closely related to the dynamics of the
underlying Ising model.

\begin{figure}[hp]
\centerline{\psfig{figure=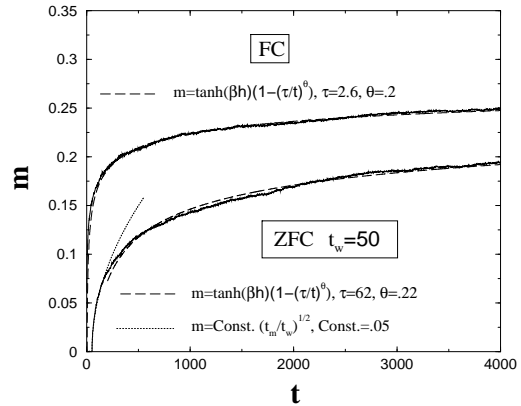,width=7.0cm,angle=0}}
\vspace{0.5cm}
\caption{Fits of the FC and ZFC data at $T/J=0.6$ with the help of Eqs. (7-9).}
\label{fig3}
\end{figure}

To understand the magnetization process  we therefore only have to consider
$c_0(t_m,t_w)$ for the Ising model.
After the first few steps, domains can be identified, and 0-sites are
on the boundary of the domains. On a square lattice, they
correspond to the diagonal portions of the boundary. Now, after a time $t_w$
the characteristic size
of the domains is of order $\sqrt{t_w}$.
Consequently the total number of domains is of order $1/t_w$, and the total
length of domain walls is proportional to $1/\sqrt{t_w}$.
When $t_m \ll t_w$ we remark two important points: (i)
the domain walls appear locally
flat, the effects of surface tension and hence curvature
are negligible and hence the domain walls diffuse
(as zero modes are the only ones generating dynamics by this stage),
a given point on a domain wall will therefore diffuse a length
$\sqrt{t_m}$; (ii)
one can  neglect the
interaction between domain walls, that is to say the coalescence of domains.
The total number of spins which
flip between $t_m$ and $t_m+t_w$ is therefore proportional to
$\sqrt{t_m} \times$
{\rm total length of domain walls}. We therefore find
that at short times $t_m$ (compared to $t_w$) one has
\begin{equation}
c_0(t_m,t_w) = Const. \sqrt{t_m\over t_w}
\end{equation}
where the constant should be independent of $t_w$.
This provides an excellent fit of the data at short times (see Fig. (\ref{fig3}).
We have checked that $Const.\simeq 0.05$ at $T=.6 J$  is indeed independent
of $t_w$.
This argument provides a
very simple explanation of the fact that the larger $t_w$,
the slower the initial increase in the magnetization.

In the case where $t_w = 0$, the above argument cannot be applied as it is
because of the initial steps, where flips occur at sites with 3 and 4
parallel neighbors as well, and because domains are not present right away.
Still the same kind of reasoning suggest that every spin which
has flipped has an average magnetization $\tanh(\beta h)$. Therefore
one has $m(t,0) = \tanh(\beta h)(1-p(t,0))$ where $p(t,0)$ is the
probability that a given spin has not flipped before time $t$. The quantity
$p(t,0)$ has received much attention in the literature
\cite{satya} and is called
the probability of persistence. Extensive numerical simulations have shown
that for large  times $p(t) \sim 1/t^{\theta}$, where $\theta$  is the
persistence exponent which has been measured as $\theta \simeq 0.2$
\cite{chate} for the
two dimensional Ising model. One therefore deduces  that
\begin{equation}
m(t,0) = \tanh(\beta h)(1-({\tau/ t})^\theta)
\end{equation}
Here $\tau$ is a microscopic time scale related to the flipping rate
of the spins. This provides and excellent fit of the data except at
very short times (see
Fig. \ref{fig3}).

In the case of $t_w$ nonzero, and for $t \gg t_w$,
one may adopt a coarse grained point of view
and cut the system up into units  of size $\sqrt t_w$. These blocks
may be regarded as effective spins of the sign of the average block
staggered magnetization.
This is related to  block persistence introduced in \cite{sire}.
Moreover, the characteristic time scale
$\tau(t_w)$ for these effective spins is proportional to the time it takes to
reverse the staggered magnetization of a given block, since
this is done by diffusion one has
 $\tau(t_w) \sim t_w$. One may now apply the reasoning of the
case $t_w=0$ to obtain
\begin{equation}
m(t,t_w) = \tanh(\beta h)(1-({\tau(t_w)/ t})^\theta)
\end{equation}
for $t \gg t_w$. This again provides a reasonable fit of the data
(see Fig. (\ref{fig3})), and we have checked that $\tau (t_w)$ is proportional
to $t_w$.

To summarize, the very slow onset of magnetization in ZFC experiments
follows scaling laws typical of ageing, and this is the origin of the
spin-glass behavior. This ageing is a consequence of the very slow
march towards equilibrium of the underlying model in which spins and
pseudo-spins are coupled into an effective Ising spins since the
physical field, which couples only to the spin and {\it not} to the
pseudo-spin, does not act as a symmetry breaking field for that model.

\section{Discussion}

While the difference between FC and ZFC measurements of the magnetization
is very typical of spin-glasses, a {\it bona fide} spin-glass has many other
characteristics which are not shared by the present model.

First of all, for an infinite system the freezing transition of a
standard spin glass is believed to be a thermodynamic transition, and
the freezing temperature is expected to be independent of
the protocol. In the present case, in order to flip, a
spin inside a domain must be able
to cross an energy barrier of $8 J$. The Arrhenius law gives therefore
a characteristic flipping time of $\tau_a \sim \exp(8J/T)$. Such activated
flips are only observed after a time of measurement of order
$\tau_a$. One may therefore define a temperature
depending on the time scale of the measurement $T_g(t_m)$, such that if
$\tau_a \gg t_m$ such activated spin flips do not occur and hence the
equilibration of the magnetization within domains is not possible.
The crossover between the two regimes is therefore  given by $T_g(t_m)/J
\sim 8/\ln(t_m)$. So, unlike the standard  thermodynamic spin glass
transition, the freezing temperature depends on $t_m$.
This estimate of $T_g$ agrees with our
Monte Carlo results. If one takes a measurement time of
$t_m = 500$, as for the curves shown in
Fig. (\ref{fig1}), one finds that  $T_g/J \sim 1.287$, close to the value of
$T_g$ shown  in Fig. (\ref{fig1}).

Another original aspect of our data with respect to standard spin
glasses is the minimum of the ZFC magnetization at a temperature
far below $T_g$. To understand this,
one may make an approximate decomposition of the magnetization
in terms of the activated component $m_a$ within the domains and the
component generated by the domain walls $m_d$. We find
$m_a(t_m, t_w) = \tanh(\beta h)(1- \exp(-\alpha(h) t_m/\tau_a))$
(solving the Glauber dynamics explicitly for a single spin
with its four neighbors fixed parallel shows that $\alpha = 1 +
\cosh(2 \beta h) + O(1/\tau_a^2)$). As $m_a$ is generated
within domains and hence in the bulk of the system (that is to say not
at domain interfaces), it is very weakly dependent on $t_w$. Another
type of activated process may occur at domain interfaces where a spin
with three neighbors antiparallel and one parallel flips. The energy
barrier for this is $4J$. This gives another characteristic time
$\tau_a^*$ with a corresponding time dependent temperature $T^*(t_m)/J
\sim 4/\ln(t_m) $. For $t_m = 500$ as in Fig. (\ref{fig1}) one finds
$T^*/J = 0.6436$ which corresponds to the minimum seen on the $ZFC$
curve.  For $T \ll T^*$, the system
behaves as if $J$ is infinite on the experimental time scale and only domain
wall diffusion occurs, the only energy involved being the external
field energy $h$. Here the dynamics is well described by
Eq. (\ref{gen}), hence the measured magnetization $m(t_m,t_w)$
increases as $T$ is lowered below $T^*$.

Both these effects are dynamic in nature and show that the behavior
we have observed is typical of the out of equilibrium dynamics
observed in  spin-glasses. This conclusion
is in fact consistent with another aspect of our data, namely the sensitivity
of the results to the dynamics used in the simulation. To perform
Monte Carlo simulations, we have made the assumption
that spins and pseudo-spins flip independently of each other. However
if one allows simultaneous flips as well, the behavior will be very
different: These moves will allow the magnetization to develop inside the
Ising domains, and the system will polarize after only a few sweeps.
In other words, the breaking of ergodicity is related to the dynamics.

Another way to check how different the present model is from standard
spin glasses is to study the temperature dependence of the non-linear
susceptibility, which is expected to diverge at the transition in
a spin-glass. Preliminary results\cite{ferrero} indicate that the
non-linear susceptibility is not singular in the present case. The absence
of another thermodynamic singularity below the Curie temperature $T_c$
is also clear from Eq. (\ref{elibre}).

However, other characteristic aspects of spin-glasses are also shared by the
present model.  For instance, the thermo-remnant magnetization
$m_{TMR}$ shows a very strong dependence on the time $t_w$ elapsed
between the quench and the switching off of the magnetic field: The
longer $t_w$, the slower the decay.
A detailed analysis of these results along the same line is in progress.

Coming back to our original purpose, namely the explanation of the low-temperature
behavior of Kagome antiferromagnets, the present model has both merits and
drawbacks. The very clear difference between FC and ZFC magnetizations
is the most interesting aspect of the results. It shows that the presence
of an extra degree of freedom which is not coupled to the external field
can indeed lead to a glassy behavior at low temperatures.
However most of the differences with standard spin-glasses are problematic
in that respect:
No experimental indication of a dependence of $T_g$ on the protocol
was reported,
and the non-linear susceptibility is indeed enhanced close to $T_g$. This is
not a final blow however. In fact, these differences with standard
spin glasses all depend on the
fact that flipping simultaneously a spin and a pseudo-spin leaves the Ising
spin unchanged. This symmetry will not be present in more realistic models
where spins and pseudo-spins are treated as Heisenberg variables, while
the underlying mechanism for the difficulty that the system will
have to magnetize
after being cooled in zero-field is still expected to apply.
More realistic models with two degrees of freedom treated
as Heisenberg spins are therefore good candidates
for effective models
to get a spin-glass behavior in disorder free magnets.
Work is in
progress along these lines.

We thank M. Ferrero and F. Becca for allowing us to quote some results
of Ref.\cite{ferrero} before publication. This work was partially
supported by the Swiss National Science Foundation under grant
number 21-63749.00.


\begin{references}
\bibitem{note}One should note that a field
coupled to both spins would break the symmetry between the phases and
allow even an infinite system to reach equilibrium in a finite time.
\end{references}
\end{document}